# MISO NOMA Downlink Beamforming Optimization with Per-Antenna Power Constraints

Yongwei Huang, Longtao Zhou

*Abstract*—Consider a multiuser downlink beamforming optimization problem for the non-orthogonal multiple access (NOMA) transmission in a multiple-input single-output (MISO) system. The total transmission power minimization problem is formulated subject to both per-antenna power constraints and quality-of-service (QoS) constraints under the NOMA principal. The problem is a non-convex quadratically constrained quadratic program, and the conventional semidefinite program (SDP) relaxation is not tight. In order to tackle the non-convex NOMA beamforming problem, we construct a second-order cone approximation for each signal-to-interference-plus-noise ratio (SINR) constraint and form an iterative algorithm, in which a sequence of second-order cone programs (SOCPs) are solved. The optimal values of the sequence of SOCPs are non-increasing, and it converges to a locally optimal value. However, our extensive simulation results show that the locally optimal value is more or less as good as the globally optimal value. In particular, we show that the SDP relaxation is tight for two-user case if one of the SINR constraints is strict (non-binding) at the optimality. Detailed simulation results are presented to demonstrate the performance gains of the NOMA downlink beamforming with per-antenna power constraints through the proposed approximate algorithm.

*Index Terms*—Non-orthogonal multiple access, downlink beamforming, per-antenna power constraint, SOCP based approximate algorithm, secure beamforming.

## I. Introduction

In a multiuser system, the effective utilization of non-orthogonal multiple access (NOMA) beamforming has been more and more popular, due to its features of improvement of spectral efficiency, user fairness, the system throughput and so on. The NOMA beamforming allows the base station (BS) to apply superposition coding (SC) using the spatial degree of freedom, and the receivers to conduct successive interference cancellation (SIC) with manageable costs, often assuming perfect channel state information known at the BS. Nowaday, the NOMA beamforming has been an important radio access technique, which is suitable for the fifth generation (5G) wireless networks [1], [2].

In this work, we deal with an NOMA beamforming optimization problem with per-antenna power constraints in setting of a multiple-input single-output (MISO) multiuser downlink transmission system. We formulate a total transmission power minimization problem subject to both per-antenna power constraints and quality-of-service (QoS) constraints under the NOMA scheme, and propose a low per-iteration complexity algorithm based on second-order cone program (SOCP) approximation. In order to proceed, let us mention some related works. In [3], multicast beamforming with SC for multiresolution broadcast with two users is considered, and the optimal beamforming vectors are obtained by minimizing the total transmission power for given target rates in a broadcast system with multiresolution transmissions where the interference is mitigated at a user using SIC. As well, the proposed multicast beamforming with SC is applied to NOMA systems as a two-stage beamforming method. A sum rate maximization problem in a MISO downlink system is studied in [4] relying on NOMA principles, and a concave-convex procedure (CCP) based iterative algorithm is developed to solve the NOMA sum rate maximization problem. In [5], the application of the concept of quasi-degradation to a multi-user MISO downlink is considered, by employing the idea of user pairing. In particular, a QoS optimization problem for two users is formulated by minimizing the total transmit power constrained by the target individual rate, and closed-form expressions for different precoding algorithms including dirty-paper coding (DPC) and zero-forcing beamforming (ZFBF) are obtained. Then the hybrid NOMA precoding algorithm is presented, which is combined with the proposed user pairing algorithm to yield practical transmission schemes. In [6], several efficient optimization algorithms for multiuser SC beamforming (SCBF) is studied to solve power minimization and rate maximization problems, and combination SCBF with ZFBF through user grouping is proposed. Optimal results are guaranteed to be found for the two user case, and it is experimentally shown that a nearly optimal performance can be achieved for the three user case with practical problem setups.

In this paper, we propose an SOCP based approximate algorithm for the MISO NOMA downlink beamforming optimization problem. The problem is formulated into the total transmission power minimization subject to the per-antenna power constraints and NOMA signal-to-interference-plus-noise rate (SINR) constraints. In the optimization problem, the number of the SINR constraints is bigger than that of those in the traditional downlink beamforming problem [7], and unlike the results in [8], the semidefinite program (SDP) relaxation is not tight [9] any more. Nevertheless, we take into account an SOCP approximation for each SINR constraint for any given initial point, and the new approximation problem is an SOCP, which can be solved efficiently. The optimal solution of the SOCP is treated as an iterative point for the next step. In the way, we obtain a sequence of SOCPs with the property that the optimal values are non-increasing. Therefore, it converges to a locally optimal solution. Our extensive simulations show that as long as the number of users is not high, the locally optimal value is very close to the optimal value of the corresponding SDP relaxation problem, which is our benchmark. In particular, when the two-user scenario is considered, we show that the SDP relaxation is tight if one of SINR inequality constraints is strict at the optimality. The approximate algorithm can be extended to solve the NOMA beamforming problem with additional secure beamforming constraints for the scenario that there are external eavesdroppers.

The paper is organized as follows. In Section II, we introduce the system model and formulate the MISO NOMA downlink beamforming problem with per-antenna power constraints. In Section III, we propose the an SOCP based approximate algorithm to solve the downlink beamforming problem, and present an extension of the approximate algorithm to solve the downlink beamforming problem with additional secure beamforming constraints. In Section IV, we present numerical examples showing the performance of the approximate algorithm in difference scenarios. Finally, the paper is concluded in Section V.

## II. System Model and Problem Formulation

Consider a wireless NOMA downlink system where a $K$-antenna base station (BS) serves $M$ single-antenna users. The transmitted

Y. Huang and L. Zhou are with School of Information Engineering, Guangdong University of Technology, University Town, Guangzhou, Guangdong 510006, China. Emails: ywhuang@gdut.edu.cn, 2111603086@mail2.gdut.edu.cn

signal by the BS is expressed as

$$\boldsymbol{x} = \sum_{m=1}^{M} s_m \boldsymbol{w}_m,$$

where $\boldsymbol{w}_m \in \mathbb{C}^K$ and $s_m$ are the downlink beamforming vector and the information symbol for user $m$ (with zero mean and unit variance), respectively. The received signal by user $m$ is given by

$$y_m = \boldsymbol{h}_m^H \boldsymbol{x} + n_m,$$

where $\boldsymbol{h}_m$ is the channel vector between the BS and user $m$, and $n_m$ is a zero-mean complex additive white Gaussian noise with variance $\sigma_m^2$ at user $m$.

In order to conduct SIC at the users, we need to build a decoding sequence, which is associated with power level of the users. We suppose that $\boldsymbol{h}_m$, $m = 1, \ldots, M$, follow the Rician channel model [10]:

$$\boldsymbol{h}_m = \sqrt{\beta_m} \left( \sqrt{\frac{\zeta}{1+\zeta}} \boldsymbol{a}(\theta_m) + \sqrt{\frac{1}{1+\zeta}} \boldsymbol{u}_m \right), \quad (1)$$

where $\boldsymbol{u}_m$ is a complex-valued Guassian random vector with zero mean and $\frac{1}{K} \boldsymbol{I}$ covariance, and

$$\boldsymbol{a}(\theta) = \frac{1}{\sqrt{K}} [1; e^{-j2\pi(d/\lambda)\sin\theta}; \ldots; e^{-j(K-1)2\pi(d/\lambda)\sin\theta}]$$

(a column vector) is the array respond vector (steering vector) for a uniform linear array (ULA) of half-wavelength spacing. Here $\theta_m$ is the angle of departure (AoD) to user $m$, and the large scaling fading factor $\beta_m$ is given by $1/(d_m)^\eta$, where $d_m$ is distance between the BS and receiver $m$, and $\eta$ is the path loss exponent (a nonnegative number). Clearly, if $\zeta = 0$, then $\boldsymbol{h}_m = \sqrt{\beta_m} \boldsymbol{u}_m$ which is a Rayleigh fading channel model (e.g., see [6]). If $\zeta = +\infty$, then $\boldsymbol{h}_m = \sqrt{\beta_m} \boldsymbol{a}(\theta_m)$, which is the scenario that the BS is equipped with a ULA with antenna spacing $d$.

Now suppose that the user set $\mathcal{S} = \{u_1, u_2, \ldots, u_M\}$ is an ordered set[1], and a user with stronger channel condition (the distance between the BS and the user is smaller) has a bigger index. Therefore, $u_m$ can decode the information for $u_n$ with $n \leq m$, under the QoS conditions

$$\min_{n \leq m \leq M} \{\mathrm{SINR}_m^n\} \geq \gamma_n, \ 1 \leq n \leq M$$

where the $\mathrm{SINR}_m^n$ is given by

$$\mathrm{SINR}_m^n = \frac{|\boldsymbol{h}_m^H \boldsymbol{w}_n|^2}{\sum_{i=n+1}^{M} |\boldsymbol{h}_m^H \boldsymbol{w}_i|^2 + \sigma_m^2}, \ n \leq m, \quad (2)$$

and

$$\gamma_n = 2^{R_n} - 1$$

with $R_n$ the achievable rate for user $u_n$.

Therefore, we consider the following NOMA downlink beamforming with per-antenna power constraints:

$$\underset{\{\boldsymbol{w}_m\}}{\text{minimize}} \quad \sum_{m=1}^{M} \boldsymbol{w}_m^H \boldsymbol{w}_m \tag{3a}$$

$$\text{subject to} \quad \boldsymbol{e}_k^H \left( \sum_{m=1}^{M} \boldsymbol{w}_m \boldsymbol{w}_m^H \right) \boldsymbol{e}_k \leq P_k, \ k = 1, \ldots, K, \tag{3b}$$

$$\min_{n \leq m \leq M} \{\mathrm{SINR}_m^n\} \geq \gamma_n, \ 1 \leq n \leq M, \tag{3c}$$

---

[1]As for how to determine an optimal decoding sequence for a general number of users is an open problem [11]. Here we assume that the order is associated with the distance between the BS and a user due to the assumption that the channel vectors $\boldsymbol{h}_m$ follow the Rician channel model (see (1)).

where $\boldsymbol{e}_k$ is the $k$-th column of the identity matrix of dimension $K$. Evidently, (3b) includes the per-antenna power constraints and $P_k$ is a given power upper bound of each per-antenna constraint. The constraints in (3c) are the user QoS conditions for a NOMA downlink system. Alternatively, we note that the following downlink beamforming problem with per-antenna power constraints can be considered:

$$\underset{\{\boldsymbol{w}_m\}, \alpha}{\text{minimize}} \quad \alpha \tag{4a}$$

$$\text{subject to} \quad \boldsymbol{e}_k^H \left( \sum_{m=1}^{M} \boldsymbol{w}_m \boldsymbol{w}_m^H \right) \boldsymbol{e}_k \leq \alpha P_k, \ k = 1, \ldots, K, \tag{4b}$$

$$\min_{n \leq m \leq M} \{\mathrm{SINR}_m^n\} \geq \gamma_n, \ 1 \leq n \leq M, \tag{4c}$$

This is an extension of the problem studied in [8]. We first focus on problem (3) and design an efficient approximate algorithm, and then show that the algorithm is applicable to (4). To determine a best $P_k$ in (3b), we can select the optimal $\alpha^\star P_k$ in (4b).

## III. AN SOCP BASED APPROXIMATE ALGORITHM FOR THE NOMA DOWNLINK BEAMFORMING PROBLEM

In this section, we establish an SOCP based approximate algorithm for (3). To this end, invoking the SINR expressions (2), the NOMA downlink beamforming problem is rewritten into a separable quadratically constrained quadratic program (QCQP):

$$\underset{\{\boldsymbol{w}_m\}}{\text{minimize}} \quad \sum_{m=1}^{M} \boldsymbol{w}_m^H \boldsymbol{w}_m \tag{5a}$$

$$\text{subject to} \quad \left( \sum_{m=1}^{M} |\boldsymbol{e}_k^H \boldsymbol{w}_m|^2 \right) \leq P_k, \ k = 1, \ldots, K, \tag{5b}$$

$$|\boldsymbol{h}_m^H \boldsymbol{w}_n|^2 \geq \gamma_n \sum_{i=n+1}^{M} |\boldsymbol{h}_m^H \boldsymbol{w}_i|^2 + \gamma_n \sigma_m^2, \tag{5c}$$

$$n \leq m \leq M, \ 1 \leq n \leq M.$$

Observe that the per-antenna power constraints are convex but the NOMA SINR constraints are non-convex. Furthermore, the total number of SINR constraints is $M(M+1)/2$, instead of $M$ as in [8]. Therefore, the traditional SDP relaxation for (5) is not tight any longer (e.g. see [9]), which indicates that we need to establish a new method without involving the SDP relaxation. Toward that, introducing auxiliary variables $\{t_{mn}\}$, problem (5) can be further recast into

$$\underset{\{\boldsymbol{w}_m\}, \{t_{mn}\}}{\text{minimize}} \quad \sum_{m=1}^{M} \boldsymbol{w}_m^H \boldsymbol{w}_m \tag{6a}$$

$$\text{subject to} \quad \text{(5b) satisfied}, \tag{6b}$$

$$|\boldsymbol{h}_m^H \boldsymbol{w}_n| \geq t_{mn} \tag{6c}$$

$$t_{mn} \geq \sqrt{\gamma_n \sum_{i=n+1}^{M} |\boldsymbol{h}_m^H \boldsymbol{w}_i|^2 + \gamma_n \sigma_m^2}, \tag{6d}$$

$$n \leq m \leq M, \ 1 \leq n \leq M.$$

It is seen that beamforming problem (6) includes the only nonconvex constraints (all others are convex):

$$|\boldsymbol{h}_m^H \boldsymbol{w}_n| \geq t_{mn}, \ n \leq m \leq M, \ 1 \leq n \leq M.$$

In order to design an algorithm for solving (6), we suppose that $\{\boldsymbol{w}_1^0, \ldots, \boldsymbol{w}_M^0\}$ is any given initial point (to be updated). Observe that

$$|\boldsymbol{h}_m^H \boldsymbol{w}_n| \geq \frac{|\boldsymbol{w}_n^{0H} \boldsymbol{h}_m \boldsymbol{h}_m^H \boldsymbol{w}_n^0|}{|\boldsymbol{h}_m^H \boldsymbol{w}_n^0|} \geq \frac{\Re(\boldsymbol{w}_n^{0H} \boldsymbol{h}_m \boldsymbol{h}_m^H \boldsymbol{w}_n^0)}{|\boldsymbol{h}_m^H \boldsymbol{w}_n^0|}. \tag{7}$$

It follows that if
$$\frac{\Re(\boldsymbol{w}_n^H \boldsymbol{h}_m \boldsymbol{h}_m^H \boldsymbol{w}_n^0)}{|\boldsymbol{h}_m^H \boldsymbol{w}_n^0|} \geq t_{mn},$$
then one has
$$|\boldsymbol{h}_m^H \boldsymbol{w}_n| \geq t_{mn}.$$
Therefore, let us consider the following beamforming problem in the SOCP form:
$$\underset{\{\boldsymbol{w}_m\},\{t_{mn}\}}{\text{minimize}} \sum_{m=1}^{M} \boldsymbol{w}_m^H \boldsymbol{w}_m \tag{8a}$$
$$\text{subject to } (5b), (6d) \text{ satisfied}, \tag{8b}$$
$$\frac{\Re(\boldsymbol{w}_n^H \boldsymbol{h}_m \boldsymbol{h}_m^H \boldsymbol{w}_n^0)}{|\boldsymbol{h}_m^H \boldsymbol{w}_n^0|} \geq t_{mn}, n \leq m \leq M, 1 \leq n \leq M. \tag{8c}$$

Observe that for the initial point $(\boldsymbol{w}_1^0, \ldots, \boldsymbol{w}_M^0)$, problem (8) is always a convex restriction of the orginal NOMA downlink beamforming problem (5).

To further proceed the analysis, let $l := 1$. Solve the SOCP (8), finding a solution $(\boldsymbol{w}_1^l, \ldots, \boldsymbol{w}_M^l; \{t_{mn}^l\})$. Again observe that
$$|\boldsymbol{h}_m^H \boldsymbol{w}_n| \geq \frac{|\boldsymbol{w}_n^H \boldsymbol{h}_m \boldsymbol{h}_m^H \boldsymbol{w}_n^l|}{|\boldsymbol{h}_m^H \boldsymbol{w}_n^l|} \geq \frac{\Re(\boldsymbol{w}_n^H \boldsymbol{h}_m \boldsymbol{h}_m^H \boldsymbol{w}_n^l)}{|\boldsymbol{h}_m^H \boldsymbol{w}_n^l|},$$
and that
$$\frac{\Re(\boldsymbol{w}_n^H \boldsymbol{h}_m \boldsymbol{h}_m^H \boldsymbol{w}_n^l)}{|\boldsymbol{h}_m^H \boldsymbol{w}_n^l|} \geq t_{mn} \text{ implies } |\boldsymbol{h}_m^H \boldsymbol{w}_n| \geq t_{mn}.$$
Then, similarly construct another convex restriction of (5) based on $(\boldsymbol{w}_1^l, \ldots, \boldsymbol{w}_M^l)$:
$$\underset{\{\boldsymbol{w}_m\},\{t_{mn}\}}{\text{minimize}} \sum_{m=1}^{M} \boldsymbol{w}_m^H \boldsymbol{w}_m \tag{9a}$$
$$\text{subject to } (5b), (6d) \text{ satisfied}, \tag{9b}$$
$$\frac{\Re(\boldsymbol{w}_n^H \boldsymbol{h}_m \boldsymbol{h}_m^H \boldsymbol{w}_n^l)}{|\boldsymbol{h}_m^H \boldsymbol{w}_n^l|} \geq t_{mn}, n \leq m \leq M, 1 \leq n \leq M. \tag{9c}$$

Then, solve (9) getting $(\boldsymbol{w}_1^{l+1}, \ldots, \boldsymbol{w}_M^{l+1}; \{t_{mn}^l\})$. Set $l := l+1$ and solve (9) again, and in the way an iterative procedure is formed. Let
$$v_l = \sum_{m=1}^{M} \|\boldsymbol{w}_m^l\|^2, l = 1, \ldots,$$
stand for the optimal values. We are about to show the important property that $\{v_l\}$ is a non-increasing sequence.

**Proposition III.1** *It holds that $v_l \geq v_{l+1}$ for $l \geq 1$.*

*Proof:* The idea is to prove that the optimal solution $(\boldsymbol{w}_1^l, \ldots, \boldsymbol{w}_M^l; \{t_{mn}^l\})$ for problem (8) with $\boldsymbol{w}_n^0$ replaced with $\boldsymbol{w}_n^{l-1}$, is feasible for (9). If that is the case, the optimal solution $(\boldsymbol{w}_1^{l+1}, \ldots, \boldsymbol{w}_M^{l+1}; \{t_{mn}^{l+1}\})$ for (9) simply has the property:
$$v_l = \sum_{m=1}^{M} (\boldsymbol{w}_m^l)^H \boldsymbol{w}_m^l \geq \sum_{m=1}^{M} (\boldsymbol{w}_m^{l+1})^H \boldsymbol{w}_m^{l+1} = v_{l+1},$$
for $l \geq 1$.

We have an immediate check whether $(\boldsymbol{w}_1^l, \ldots, \boldsymbol{w}_M^l; \{t_{mn}^l\})$ is feasible for (9). Since it is optimal for problem (8) with $\boldsymbol{w}_n^0$ replaced with $\boldsymbol{w}_n^{l-1}$, then it is feasible too and thus, (8c) are satisfied:
$$\frac{\Re((\boldsymbol{w}_n^l)^H \boldsymbol{h}_m \boldsymbol{h}_m^H \boldsymbol{w}_n^{l-1})}{|\boldsymbol{h}_m^H \boldsymbol{w}_n^{l-1}|} \geq t_{mn}^l,$$
which imply that
$$|\boldsymbol{h}_m^H \boldsymbol{w}_n^l| \geq t_{mn}^l. \tag{10}$$

(5b) and (6d) are also fulfilled:
$$\left( \sum_{m=1}^{M} |\boldsymbol{e}_k^H \boldsymbol{w}_m^l|^2 \right) \leq P_k, k = 1, \ldots, K, \tag{11}$$
and
$$t_{mn}^l \geq \sqrt{\gamma_n \sum_{i=n+1}^{M} |\boldsymbol{h}_m^H \boldsymbol{w}_i^l|^2 + \gamma_n \sigma_m^2}, \tag{12}$$

for $n \leq m \leq M, 1 \leq n \leq M$, respectively. It follows from (10)-(12) that $(\boldsymbol{w}_1^l, \ldots, \boldsymbol{w}_M^l; \{t_{mn}^l\})$ is feasible for (9). ∎

Remark that problems (9) for $l = 1, 2, \ldots$, are convex restrictions of (5) and the optimal values are non-increasing, namely, $v_1 \geq v_2 \geq \ldots$. Therefore $(\boldsymbol{w}_1^l, \ldots, \boldsymbol{w}_M^l)$ converges to a locally minimal point for (5) (our extensive numerical simulations show that it goes to the globally minimal point when the number of users is not too high).

Note that we can remove the auxiliary variables, and reformulate (9) equivalently into a compact form
$$\underset{\{\boldsymbol{w}_m\}}{\text{minimize}} \sum_{m=1}^{M} \boldsymbol{w}_m^H \boldsymbol{w}_m \tag{13a}$$
$$\text{subject to } (5b) \text{ satisfied}, \tag{13b}$$
$$\frac{\Re(\boldsymbol{w}_n^H \boldsymbol{h}_m \boldsymbol{h}_m^H \boldsymbol{w}_n^l)}{|\boldsymbol{h}_m^H \boldsymbol{w}_n^l|} \geq \sqrt{\gamma_n \sum_{i=n+1}^{M} |\boldsymbol{h}_m^H \boldsymbol{w}_i|^2 + \gamma_n \sigma_m^2}$$
$$\tag{13c}$$
$$n \leq m \leq M, 1 \leq n \leq M$$

Therefore, we summarize an approximate algorithm based on the SOCP restrictions as in Algorithm 1.

---

**Algorithm 1** An Approximate algorithm for (5)

**Input:** $\{\boldsymbol{h}_m\}$, $\{\sigma_m\}$, $\{\gamma_n\}$, $\{P_k\}$, $M$, $K$, $\zeta$, $\eta$, $\xi$;
**Output:** A solution $\{\boldsymbol{w}_m^\star\}$ for problem (5);
1: Suppose that $\{\boldsymbol{w}_m^0\}$ is an initial point; set $l = 0$ and $v_0 = \sum_{m=1}^{M} \|\boldsymbol{w}_m^0\|^2$ (a large number);
2: **do**
3:     solve SOCP (13), obtaining optimal solution $\{\boldsymbol{w}_m^{l+1}\}$ and optimal value $v_{l+1}$;
4:     $l := l + 1$;
5: **until** $v_{l-1} - v_l \leq \xi$

---

As for the computational complexity, it mainly includes solving SOCP (13) in every iteration. Let us check the worst-case computational complexity for (13) based on [12, page 309]. There are $K + 1 + M(M+1)/2$ constraints in the SOCP, the length of the optimization (real-valued) variables are $2NM + 1$, and $(K+1)M^2 + \sum_{i=1}^{M(M+1)/2} k_i^2$ is equal to $(K+1)M^2 + \sum_{m=1}^{M} M^3 = (K+1)M^2 + (M(M+1)/2)^2$. Therefore, the total complexity is of order $O((M(M+1)/2 + K + 2)^{1/2}(2NM+1)((2NM+1)^2 + (K+1)(M^2+1) + (M(M+1))/2 + (M^2(M+1)^2)/4)$, which is approximately equal to $O(\sqrt{K+M^2}NM^3(8N^2 + 2K + M^2/2))$. Recall that the complexity is the worst-cast theoretical value, and it is not real computational cost.

We remark that from optimization point of view, problem (5) with $\boldsymbol{h}_m \boldsymbol{h}_m^H$ replaced with $\boldsymbol{H}_m$ (general rank) can be solved similarly by using the approximate algorithm.

As for how solve (4), we construct the following SOCP restriction problem similarly:
$$\underset{\{\boldsymbol{w}_m\},\{t_{mn}\},\alpha}{\text{minimize}} \alpha \tag{14a}$$
$$\text{subject to } (4b), (6d), (9c) \text{ satisfied}, \tag{14b}$$

for $l = 0, 1, \ldots$. Like the proof of Proposition III.1, it is not hard to show that the feasible set of SOCP (14) with $l$ is always contained in that of the SOCP with $l+1$ for $l = 1, 2, \ldots$. Therefore, we have $\alpha_l \geq \alpha_{l+1}$ for all $l \geq 1$. Algorithm 1 is applicable to solve (4), simply by changing SOCP (13) in step 3 into (14) with both constraints (6d) and (9c) replaced by (13c).

*A. Two-User Case*

If there are two users in the NOMA downlink transmission system, then problem (5) is specified to

$$\underset{\{\boldsymbol{w}_1, \boldsymbol{w}_2\}}{\text{minimize}} \quad \boldsymbol{w}_1^H \boldsymbol{w}_1 + \boldsymbol{w}_2^H \boldsymbol{w}_2 \tag{15a}$$

$$\text{subject to} \quad |\boldsymbol{e}_k^H \boldsymbol{w}_1|^2 + |\boldsymbol{e}_k^H \boldsymbol{w}_2|^2 \leq P_k, \; k = 1, \ldots, K \tag{15b}$$

$$\boldsymbol{w}_1^H \boldsymbol{h}_1 \boldsymbol{h}_1^H \boldsymbol{w}_1 - \gamma_1 \boldsymbol{w}_2^H \boldsymbol{h}_1 \boldsymbol{h}_1^H \boldsymbol{w}_2 \geq \gamma_1 \sigma_1^2 \tag{15c}$$

$$\boldsymbol{w}_1^H \boldsymbol{h}_2 \boldsymbol{h}_2^H \boldsymbol{w}_1 - \gamma_1 \boldsymbol{w}_2^H \boldsymbol{h}_2 \boldsymbol{h}_2^H \boldsymbol{w}_2 \geq \gamma_1 \sigma_2^2 \tag{15d}$$

$$\boldsymbol{w}_2^H \boldsymbol{h}_2 \boldsymbol{h}_2^H \boldsymbol{w}_2 \geq \gamma_2 \sigma_2^2, \tag{15e}$$

for which the traditional SDP relaxation is

$$\underset{\{\boldsymbol{w}_1, \boldsymbol{w}_2\}}{\text{minimize}} \quad \text{tr}\, \boldsymbol{W}_1 + \text{tr}\, \boldsymbol{W}_2 \tag{16a}$$

$$\text{subject to} \quad \text{tr}\,(\boldsymbol{e}_k \boldsymbol{e}_k^H (\boldsymbol{W}_1 + \boldsymbol{W}_2)) \leq P_k, \; k = 1, \ldots, K \tag{16b}$$

$$\text{tr}\,(\boldsymbol{h}_1 \boldsymbol{h}_1^H (\boldsymbol{W}_1 - \gamma_1 \boldsymbol{W}_2^H)) \geq \gamma_1 \sigma_1^2 \tag{16c}$$

$$\text{tr}\,(\boldsymbol{h}_2 \boldsymbol{h}_2^H (\boldsymbol{W}_1 - \gamma_1 \boldsymbol{W}_2^H)) \geq \gamma_1 \sigma_2^2 \tag{16d}$$

$$\text{tr}\,(\boldsymbol{h}_2 \boldsymbol{h}_2^H \boldsymbol{W}_2) \geq \gamma_2 \sigma_2^2 \tag{16e}$$

$$\boldsymbol{W}_1 \succeq \boldsymbol{0}, \; \boldsymbol{W}_2 \succeq \boldsymbol{0}. \tag{16f}$$

The dual is the following maximization problem:

$$\underset{\{y_m\}, \{x_k\}}{\text{maximize}} \quad y_1 \gamma_1 \sigma_1^2 + y_2 \gamma_1 \sigma_2^2 + y_3 \gamma_2 \sigma_2^2 - \sum_{k=1}^{K} x_k P_k \tag{17a}$$

$$\text{subject to} \quad \boldsymbol{I} - y_1 \boldsymbol{h}_1 \boldsymbol{h}_1^H - y_2 \boldsymbol{h}_2 \boldsymbol{h}_2^H + \sum_{k=1}^{K} x_k \boldsymbol{e}_k \boldsymbol{e}_k^H \succeq \boldsymbol{0} \tag{17b}$$

$$\boldsymbol{I} + y_1 \gamma_1 \boldsymbol{h}_1 \boldsymbol{h}_1^H + y_2 \gamma_1 \boldsymbol{h}_2 \boldsymbol{h}_2^H - y_3 \boldsymbol{h}_2 \boldsymbol{h}_2^H +$$
$$\sum_{k=1}^{K} x_k \boldsymbol{e}_k \boldsymbol{e}_k^H \succeq \boldsymbol{0} \tag{17c}$$

$$y_1 \geq 0, y_2 \geq 0, y_3 \geq 0, x_k \geq 0, k = 1, \ldots, K. \tag{17d}$$

**Theorem III.2** *Suppose that $(\boldsymbol{W}_1^\star, \boldsymbol{W}_2^\star)$ is an optimal solution for SDP (16) and $\boldsymbol{W}_i^\star \neq \boldsymbol{0}$ for $i = 1, 2$. Then $\boldsymbol{W}_2^\star$ is of rank one. If either $\text{tr}\,(\boldsymbol{h}_1 \boldsymbol{h}_1^H (\boldsymbol{W}_1^\star - \gamma_1 \boldsymbol{W}_2^{\star H})) > \gamma_1 \sigma_1^2$ or $\text{tr}\,(\boldsymbol{h}_2 \boldsymbol{h}_2^H (\boldsymbol{W}_1^\star - \gamma_1 \boldsymbol{W}_2^{\star H})) > \gamma_1 \sigma_2^2$, then $\boldsymbol{W}_1^\star$ is a rank-one matrix.*

The proof is immediate if one verifies the complementary conditions (i.e., the complementary slackness in the KKT optimality conditions). We hence omit it.

Observe that at the optimality, at least one of two inequality constraints (16c)-(16d) holds with equality, and the inequality constraint (16e) becomes equality (active or binding).

*B. Extensions*

In addition to the per-antenna power constraints, let us consider a scenario that there are external eavesdroppers in the NOMA downlink transmission. To control the transmission energy around the eavesdroppers as low as possible, we introduce the following secure beamforming constraint:

$$\sum_{m=1}^{M} |\boldsymbol{f}_r^H \boldsymbol{w}_m|^2 = \boldsymbol{f}_r^H \left( \sum_{m=1}^{M} \boldsymbol{w}_m \boldsymbol{w}_m^H \right) \boldsymbol{f}_r \leq \delta_r, \; r = 1, \ldots, R \tag{18}$$

where $\boldsymbol{f}_r$ is the channel vector between the BS and eavesdropper $r$, and $\delta_r$ is the upper bound for transmission energy around eavesdropper $r$ (the bound is the lower, the better). Therefore, the new NOMA downlink beamforming problem is expressed as:

$$\underset{\{\boldsymbol{w}_m\}}{\text{minimize}} \quad \sum_{m=1}^{M} \boldsymbol{w}_m^H \boldsymbol{w}_m \tag{19a}$$

$$\text{subject to} \quad (5b), (5c), (18) \text{ satisfied}. \tag{19b}$$

Evidently, the problem can be approximated by the following SOCP:

$$\underset{\{\boldsymbol{w}_m\}}{\text{minimize}} \quad \sum_{m=1}^{M} \boldsymbol{w}_m^H \boldsymbol{w}_m \tag{20a}$$

$$\text{subject to} \quad (5b), (13c), (18) \text{ satisfied}. \tag{20b}$$

In a similar way, to solve the downlink beamforming problem (19) with the secure beamforming constraints (18), we need only to invoke the approximate algorithm changing (13) in step 3 into SOCP (20).

## IV. SIMULATIONS RESULTS

In this section, we provide simulation results demonstrating the performance of the proposed approximate algorithm solving NOMA downlink beamforming optimization problems with per-antenna power constraints. Let us consider a NOMA downlink system where the BS is equipped with $K = 8$ antennas, and the receiver noise power $\sigma_m^2 = 1$ W (Watt) for all $m$.

*Example 1.* Suppose that the BS serves $M = 4$ users (as in problem (5)). Assume that the Rayleigh fading channel $\boldsymbol{h}_m$ follows $\mathcal{N}(\boldsymbol{0}, 1/((d_m)^\eta K)\boldsymbol{I})$ (i.e. (1) with $\zeta = 0$), where $d_m$ is the distance between the BS and user $m$, and $\eta$ is the path loss exponent. We set $d_M = d_4 = 2$ m (meter), $d_3 = 5$ m, $d_2 = 10$ m, $d_1 = 15$ m, and $\eta = 2$. The power upper bound for each antenna is $P_k = 6$ W, $k = 1, \ldots, K$. The SINR targets $\gamma_4 = 1$, $\gamma_3 = 0.5$, $\gamma_2 = 0.1$, $\gamma_1 = 0.05$.

In Fig. 1, we have plotted the optimal transmission power on antenna 1 in Fig. 1 (a) for a set of 100 channel realizations (where the feasible sets of problem (5) without the per-antenna power constraints are nonempty), and in Fig. 1 (b) for another set of 100 channel realizations (where the feasible sets of problem (5) are nonempty). Both problem (5) without the per-antenna power constraints and problem (5) are solved by Algorithm 1 (rather than by solving its SDP relaxation) for every feasible channel realization. We observe in Fig. 1 (a) that without the per-antenna power constraints, there are many realizations for which the optimal transmission power on antenna 1 is over per-antenna power upper bound $P_1$, while in Fig. 1 (b) that in all realizations the transmission power on antenna 1 is equal to or below $P_1$, which is in line with the condition that the per-antenna power constraints are enforced.

*Example 2.* Suppose that the BS serves a set of six users $\{u_1, u_2, u_3, u_4, u_5, u_6\}$, which is an ordered set representing the decoding sequence with bigger index user having stronger channel. Assume that $\boldsymbol{h}_m$ follows the channel model (1) with $\zeta = 0$ (i.e. Rayleigh fading channel). The distances between the BS and the users are $(d_1, d_2, d_3, d_4, d_5, d_6) = (6, 5, 4, 3, 2, 1)$. The path loss exponent $\eta = 1$. In addition, we assume that $\gamma_1 = \cdots = \gamma_6 = 1$. The power in dB is the ratio of the power in W over the noise power in W, so that 1 corresponds to 0 dB. When we say that the BS serves four (two) users, it refers to the users $\{u_1, u_2, u_3, u_4\}$ ($\{u_1, u_2\}$). The power upper bound $P_k = 10\alpha^\star$ dB for each antenna, where $\alpha^\star$ is the optimal value for (4) with the same settings as stated in this example. Every data point in the figure is the average over 100 channel realizations.

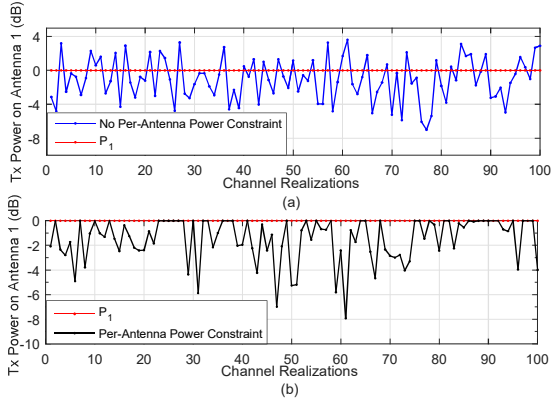

Fig. 1. The optimal transmission power on antenna 1 versus channel realizations; (a) NOMA beamforming problem (5) excluding the per-antenna power constraints; (b) NOMA beamforming problem (5). The transmission power on antenna 1 is divided by $P_1 = 6W$ so that 1 corresponds to 0 dB (thus the power upper bound $P_1$ of antenna 1 is 0 dB).

Fig. 2 demonstrates the optimal total transmission power versus the number of transmit antennas. As can be seen, less transmission power is required when the BS has more transmit antennas. This is reasonable since the optimization search space of (5) is larger (the minimal value is smaller) when the number of transmit antennas is bigger. Furthermore, more transmission power is needed when the number of users is increasing. This is also sound since more transmission power is required in order to serve additional users.

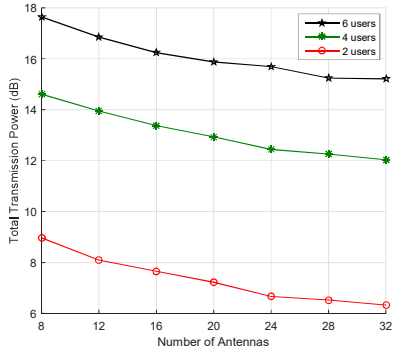

Fig. 2. The optimal total transmission power versus number of transmit antennas, with $\zeta = 0$, $\gamma = 1$, and $\eta = 1.0$.

*Example 3.* We consider a scenario that up to four users are served. We assume that the channel vector $\boldsymbol{h}_m$ follows the model (1) with $\zeta = 10$ and $\eta = 1$, and the SINR thresholds are equal to each other, i.e. $\gamma_m = \gamma$, $\forall m$. The number of transmit antennas $K = 20$. The distances $(d_1, d_2, d_3, d_4) = (6, 5, 4, 3)$, and the AoDs in (1) are $(\theta_1, \theta_2, \theta_3, \theta_4) = (30°, 40°, 50°, 50°)$. When saying three (two) users served by the BS, it means that $\{u_1, u_2, u_3\}$ ($\{u_1, u_2\}$) are selected. The power upper bound for each antenna $P_k$ is the same as that in Example 2. Other settings follow Example 2.

Fig. 3 examines how different the optimal values obtained by the SOCP based approximation and by the SDP relaxation are. This figure displays the total transmission power versus the common SINR threshold $\gamma$. We observe in Fig. 3 that the difference between the two optimal values is marginal, which means that the approximation is very closed to the globally optimal value for the NOMA downlink beamforming problem.

## V. Conclusion

In this paper, the MISO NOMA downlink beamforming optimization problem with per-antenna power constraints has been studied.

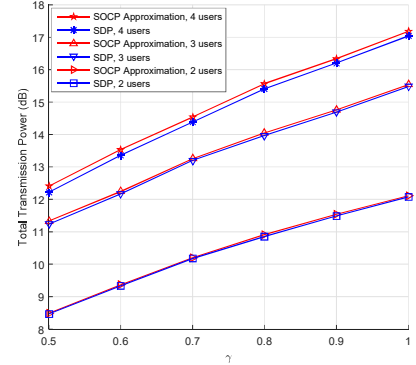

Fig. 3. The optimal total transmission power versus SINR threshold $\gamma$, with $\zeta = 10$, $\eta = 1$, and $K = 20$.

An SOCP based approximate algorithm has been proposed for the downlink beamforming problem. In words, a sequence of SOCPs are solved with the property that the feasible set of SOCP in the $l$-th iteration is always included in that of SOCP in the $(l + 1)$-th iteration. This gives the sequence of SOCPs with non-increasing optimal values, which means that the approximation converges to a locally optimal value. In particular, we have proved that the SDP relaxation for the NOMA downlink beamforming problem with two users is tight if one of the NOMA SINR inequality constraints is strict at the optimality. We have shown that the NOMA downlink beamforming problem with both per-antenna power constraints and secure beamforming constraints can be solved similarly by invoking the proposed approximate algorithm when the scenario with external eavesdroppers is considered. The performance of the proposed beamforming designs has been demonstrated by simulations.